\documentclass[12pt]{article} 
\usepackage{graphicx}
\usepackage{latexsym}

\begin{document}
\bibliographystyle{plain}

\title{A Two-dimensional Model of Shear-flow Transition}
\author{Norman R Lebovitz\\Department of Mathematics\\The University of
Chicago\\5734 S University Ave, Chicago IL 60637,
USA\\email:norman@math.uchicago.edu}

\date{June 11, 2010}
\maketitle
\abstract{We explore a two-dimensional dynamical system modeling
transition in shear flows to try to understand the nature of an 'edge'
state.  The latter is an invariant set in phase space separating the
basin of attraction $B$ of the laminar state into two parts
distinguished from one another by the nature of relaminarizing
orbits. The model is parametrized by $R$, a stand-in for Reynolds
number. The origin is a stable equilibrium point for all values of $R$
and represents the laminar flow. The system possesses four critical
parameter values at which qualitative changes take place,
$R_{sn},R_h,R_{bh}$ and $R_\infty$.  The origin is globally stable if
$R<R_{sn}$ but for $R>R_{sn}$ has two further equilibrium points,
$X_{lb}$ and $X_{ub}$. Of these $X_{lb}$ is unstable for all values of
$R>R_{sn}$ whereas $X_{ub}$ is stable for $R<R_{bh}$ and therefore
possesses its own basin of attraction $D$. At $R=R_h$ a homoclinic
bifurcation takes place with the simultaneous formation of a
homoclinic loop and an edge state. 

For $R_h < R < R_{bh}$ the edge state, which is the stable manifold
of $X_{lb}$, forms part of $\partial B$, the boundary of the basin of
attraction of the origin. The other part of $\partial B$ is
a periodic orbit $P$ bounding $D$. $P$ and $D$ shrink with increasing $R$. 
At
$R=R_{bh}$ there is a 'backwards Hopf' bifurcation at which $X_{ub}$
loses its stability and $P$ and $D$ disappear. For
$R_{bh}<R<R_\infty$ the edge is 'pure' in the sense that it is the
only phase space structure that lies outside the basin of attraction
of the origin. As $R$ increases the point $X_{lb}$ recedes to
progressively greater distances, with a singularity at $R=R_\infty$
where it becomes infinite.  For $R>R_\infty$ $X_{lb}$ has reappeared,
the edge state has disappeared, and the geometrical structure favors
permanent transition from the laminar state, increasingly so for
increasing values of $R$.}
\section{Introduction}\label{intro}
Transition to turbulence in shear flows occurs as the relevant
parameter, the Reynolds number $R$, increases beyond a critical
value. This transition differs in important respects from the onset of
instability in other familiar problems of hydrodynamics and indeed of
other familiar problems in applied mathematics. One difference of long
standing is that the transition takes place while the unperturbed,
laminar flow remains asymptotically stable
(\cite{eck08},\cite{sgreview}). A related difference is that the
transition is not sharp: the value of the critical Reynolds number for
onset depends on the size and nature of the perturbation.

Another difference is that the transition may not be permanent. For
a range of $R$ values, an apparently complex motion occurs
for a while but is then followed by relaminarization. This regime of a return
to the laminar state after a complex motion has been found in
numerical calculations and related to the occurrence of an
'edge' state (\cite{eck08},\cite{awh}). Sometimes the complex motion has a chaotic
character and one refers to the 'edge of chaos.' These features have
been found both in the behavior of low-dimensional models of
shear flows and in numerical treatments of the Navier-Stokes
equations. In these theoretical treatments the edge seems to be a
codimension-one surface in phase space separating relaminarizing
orbits of two different types: orbits of one type relaminarize quickly,
whereas those of other type relaminarize more slowly and follow a 
more complicated trajectory than orbits of the first type (cf
\cite{skufca}). An alternative picture in which the edge is a {\em pair} of
surfaces very close together was proposed in \cite{nl} but seems
implausible in light of the conclusions of the present paper (see \S
\ref{conclusions} below).

The object of the present note is to exhibit a two-dimensional
model in which some of the features described above --
particularly the edge state --  emerge in a
transparent manner. The choice of model is in some respects arbitrary,
and the fact that it is two-dimensional excludes a variety of
behaviors that are possible in the Navier-Stokes equations. It may nevertheless be useful in
supplying an internally consistent template for the behavior of such systems. In
our model the laminar state is represented by the origin of
coordinates and 'relaminarization' of an orbit means that it tends to
the origin as $t \to + \infty$. Before we turn to this model (\S \ref{hydrotype}), a word about the
basin of attraction.

\section{The basin of attraction}\label{simple}

The basin of attraction $B$ of an asymptotically stable equilibrium
point $q$ is the set of initial-data points whose orbits tend to $q$
as $t \to +\infty$. By the boundary of $B$, $\partial
B$, we mean here the set of points $p$ whose every neighborhood
contains both points that are in $B$ and points that are not in
$B$. There are simple examples for which $\partial B$ separates phase
space into two regions, $B$ and a complementary region $D$ in which
{\em no}
orbit tends to $q$. It is worthwhile emphasizing that the definition of
$\partial B$ does not require this simple picture to hold.

One simple example is
\begin{equation}\label{naive} \dot{x}=-x(1-x^2-y^2) -y,\; \dot{y}=x -y(1-x^2-y^2),
\end{equation}
which becomes $\dot{r}=-r(1-r^2),\; \dot{\theta}=1$ in polar coordinates. There is a stable equilibrium point
at the origin ($q=0$). Orbits for which $r(0)<1$
relaminarize and the basin of attraction is
precisely the set $r<1$. Orbits for which $r(0) \ge 1$ never
relaminarize. 

The 'edge' picture of
Eckhardt et al (cf \cite{skufca}) is more complex. It seems to imply a near-global
stability of the laminar point in the sense that virtually all orbits
relaminarize. There must be
exceptions -- invariant sets like equilibrium points or periodic
orbits -- and the edge
itself appears to be an invariant set. But these exceptions
are of lower dimension then the ambient space so, with probability one,
all orbits relaminarize. This picture, which is arrived at through
numerical computations, could of course be a consequence of incomplete
numerical
sampling: there could be undiscovered, open, invariant sets somewhere in phase space. 

The simple system (\ref{naive}) above, which has no edge state, is not
designed to model shear
flows. We now examine a two-dimensional system that is designed to model shear flows,
and does have an edge state.

\section{Systems of shear-flow type}\label{hydrotype}

Finite-dimensional systems that mimic shear flows possess the
following characteristics: they have only linear and quadratic terms,
the linear terms feature a non-normal matrix and the nonlinear terms
conserve energy (cf \cite{nl}). We consider the following family of two-dimensional systems of
this kind:
\begin{eqnarray}\label{gen2dsys1}
&& \dot{x}_1=-\delta x_1+  x_2 + b _1 x_1x_2 -b _2 x_2^2,\\
&& \dot{x}_2= -\delta x_2 - b _1 x_1^2 +b _2 x_1x_2.
\label{gen2dsys2}
\end{eqnarray}
The shear-flow problem is characterized by a Reynolds number $R$ and
this appears in these equations through the relation $\delta =1/R$.
The real parameters $b_1,b_2$ could be chosen arbitrarily, but in the numerical examples
below we have chosen $b_1=1, b_2=3$. The laminar flow is represented
by the equilibrium solution $(x_1,x_2)=(0,0)=O$.

\begin{figure}[bt]
\centering \includegraphics[width=4.0in, height=2.0in]{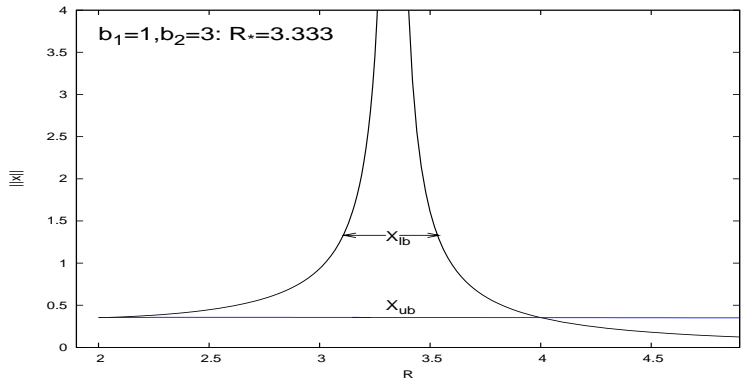}
\caption{\small  The abscissa $\|x\|=0$ represents the stable, laminar
point. The apparent cusp at the saddle-node point $R=2$ is an
artifact of the special choice  $b_1=1$ and $b_2=3$ and applies only
to the norm: the individual components $x_1$ and $x_2$ have a smooth
(parabolic) behavior there. On the other hand singular behavior as $R
\to 10/3$ (labeled $R_*$ in the figure) is common to the cases when $b_1$ and $b_2$ have the same sign.}\label{norm2fig}
\end{figure}

Further equilibrium solutions of the system
(\ref{gen2dsys1}) can be expressed as follows:
\begin{equation}\label{equilrels}
x_1=\frac{C\delta}{Cb_2-b_1} \mbox{ and } x_2=C x_1\end{equation}
where
\begin{equation}\label{Ceq}
C=\left(1 \pm \sqrt{1-4\delta^2}\right)/2\delta = \left(R \pm
\sqrt{R^2-4}\right)/2 \end{equation}
Equation (\ref{Ceq}) shows that two real equilibrium points other
than the origin $O$ occur if and only if $R>2$. This identifies the
'saddle-node' critical value as $R_{sn}=2$ in the present case. It is
not difficult to show that the origin is asymptotically stable if $R <
2$. For $R>2$ we'll denote by $X_{lb}$ the equilibrium point obtained
when the minus sign is chosen in equation (\ref{Ceq}) and by
$X_{ub}$ the point obtained when the plus sign is chosen. For the
choices $b_1=1$ and $b_2=3$ we get Figure \ref{norm2fig}, showing the
norm of an equilibrium solution plotted against $R$.

The stability of an equilibrium solution is determined from
consideration of the eigenvalues of the Jacobian matrix
\begin{equation}\label{Df}
Df=\left(\begin{array}{cc} -\delta + b_1 x_2 & 1+b_1x_1 - 2
b_2x_2\\ -2b_1x_1 + b_2x_2& -\delta + b_2x_1
\end{array}\right).
\end{equation}

\noindent Two critical values of $R$ appear in Figure \ref{norm2fig}: the
saddle-node value $R_{sn}$ and a value $R_\infty$ at which $X_{lb}$ fails
to exist (tends to infinite distance as $R \to R_\infty$). It is clear, however
that there may be further critical values of $R$ at which
the phase portraits undergo qualitative changes, and these will indeed play a
role below. Some of the results of the present and following section
can be obtained analytically using the formulas of this section.
However, to understand the geometry of phase space, there seems to be no satisfactory alternative to numerical
investigations of the basin boundary, even in this simplest of cases. We
now turn to this.

\section{Numerical examples}\label{numerical}
The objective is to understand the geometry of phase space and how it
changes with $R$.
We approach this by constructing, for a representative sequence of $R$ values, invariant curves that limit the behavior
of orbits to certain regions of phase space. The starting point in all
of the cases considered below is the construction of the stable and
unstable manifolds of the unstable equilibrium point $X_{lb}$.

\begin{figure}
\centering \includegraphics[width=3.5in, height=2.5in]{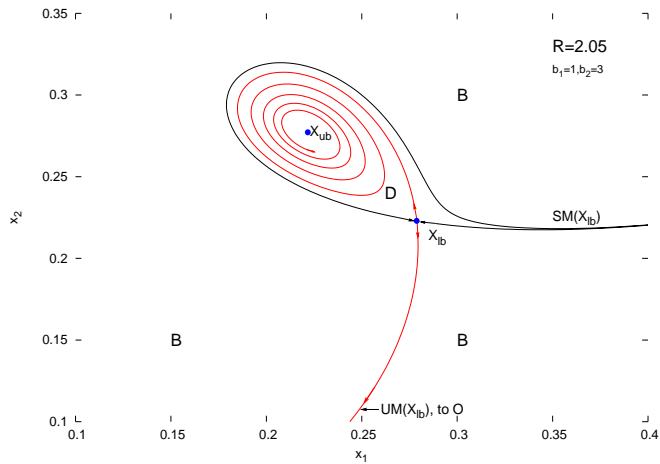}
\caption{\small For a value of $R$ just greater than the first
critical value $R=R_{sn}=2$, the phase
portrait is a familiar one in which the boundary of the basin of
attraction of $O$ is likewise the boundary of the basin of attraction
of a second, stable equilibrium point $X_{ub}$.  }\label{bb_2.05_1_3fig}
\end{figure}

\subsection{A 'simple' basin}\label{simplebasin}
For Figure \ref{bb_2.05_1_3fig} we choose a value 
$R=2.05$, just slightly greater than the saddle-node critical value. 
The equilibrium point $X_{ub}$ is stable and has its own basin of
attraction, indicated by $D$ in the figure. The remainder of the phase
space lies in $B$, the basin of attraction of the origin. The
equilibrium point $X_{lb}$ has one stable and one
unstable direction. Its stable manifold, $SM(X_{lb})$, forms the common
boundary of $D$ and $B$. This picture is similar to that of the simple
basin of attraction of \S \ref{naive} in that the basin boundary
divides phase space into the sharply distinct regions $B$, where all
orbits relaminarize, and $D$ where no orbits relaminarize. It differs
in details: the part of the domain $D$ that lies to the right of $X_{lb}$
becomes extremely thin, the two arcs of $SM(X_{lb})$ coming
progressively closer to coincidence for progressively larger values of $x_1$. It differs also in that the
basin boundary is unbounded. Before considering larger values of $R$ we we make some observations 
about the system near the saddle-node value. 

First, consider linearizing the 
system (\ref{gen2dsys1}) about $X_{sn}$, the saddle node point, for 
$R=R_{sn}$. The matrix of linearization (equation \ref{Df}) is 
\[ \left(\begin{array}{cc}-1/4&-1/4\\1/4&1/4\end{array}\right) \approx
 \left(\begin{array}{cc}0&0\\1&0\end{array}\right) \]
under similarity transformation. If we were to unfold the singularity
at that point, the unfolding would be of Takens-Bogdanov type. Of
course we have a system with prescribed nonlinear terms so we do not
undertake this unfolding but it is interesting to note that in a far more faithful representation of
the pipe-flow problem (see \cite{me}) one is led to this unfolding at
the corresponding value $R_{sn}$.

Next note that, despite the unboundedness of $D$, it has a small
area. A perturbation of the laminar flow has only a small probability
of lying in $D$ and therefore a high probability of
relaminarizing. 

Finally we note that a hint of 'edge' behavior exists in this
diagram: for values of $x_1$ exceeding (say) $0.35$, two 
initial points near $SM$, one just above the upper arc and the other
just below the lower arc, lead to quite different trajectories,
although both end up at $O$.

\subsection{Homoclinic bifurcation}\label{homoclinic}
\begin{figure}[h]
\centering \includegraphics[width=3.5in, height=2.5in]{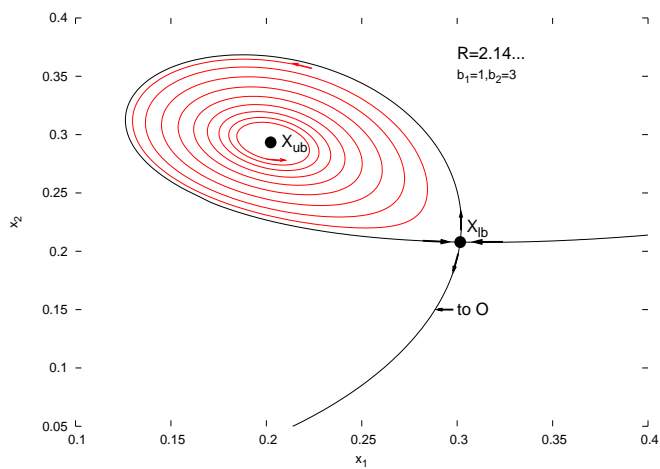}
\caption{\small At the second critical value $R=R_h = 2.14\ldots$, there
is a homoclinic bifurcation of phase portraits,
depicted here and producing an edge.}\label{hmclncfig}
\end{figure}

As $R$ increases further the region $D$ begins to increase in size but
a new critical value of $R=R_h \approx 2.14$ occurs where a homoclinic bifurcation
takes place: the two branches of $SM(X_{lb})$ to the right of $X_{lb}$
seen in Figure \ref{bb_2.05_1_3fig} coalesce into a single orbit, and $D$ is now bounded by a homoclinic loop. This is depicted in Figure
\ref{hmclncfig}. The part of the stable manifold
of $X_{lb}$ to the right of this point has now taken the form of an
edge as described elsewhere (\cite{skufca}). 

\begin{figure}
\centering \includegraphics[width=3.5in, height=2.5in]{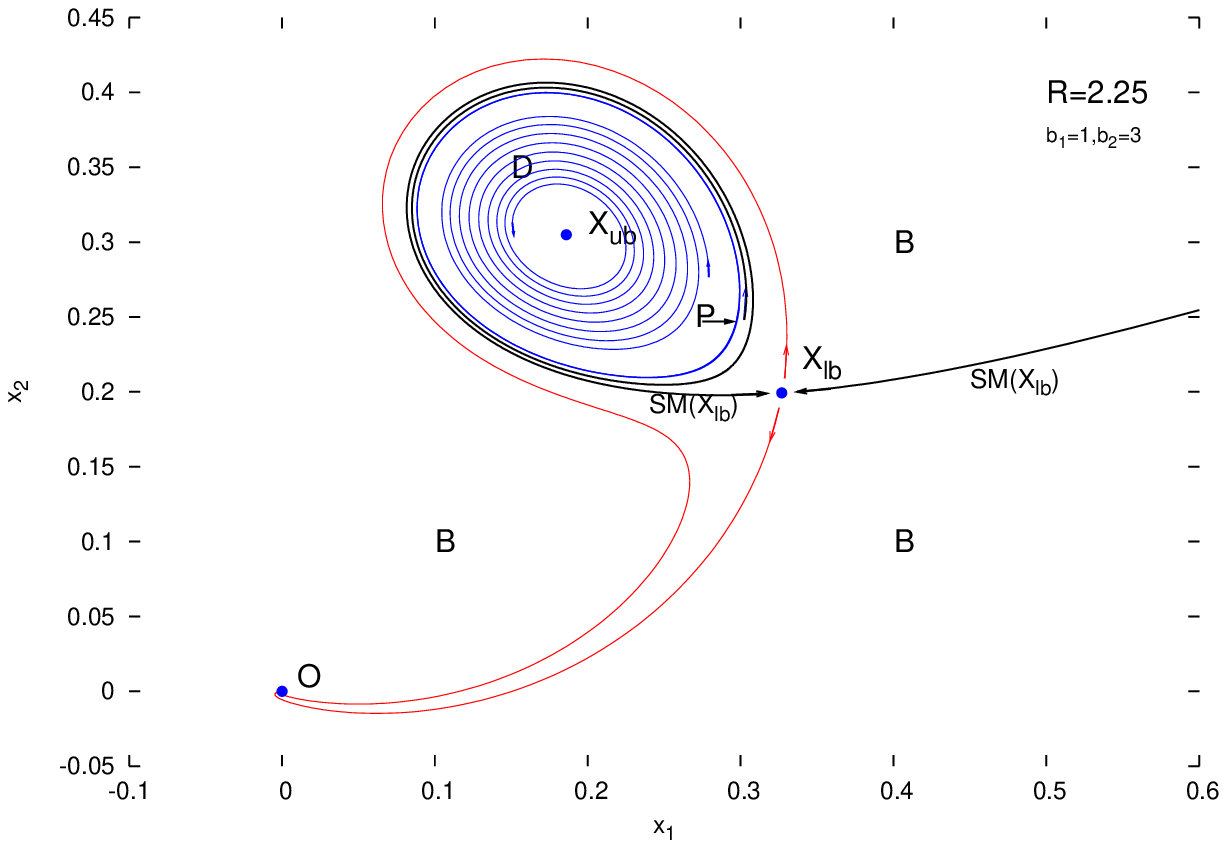}
\caption{\small For a slightly larger value of $R$ than that shown in
Figure \ref{hmclncfig}, the phase
portrait has changed qualitatively.  No longer like the familiar picture,
there is now an edge, agreeing with the stable manifold of
$X_{lb}$. The basin of
attraction of $X_{ub}$ is now bounded; its boundary is the periodic
orbit $P$.}\label{bb_2.25_1_3fig}
\end{figure}

Increasing $R$ beyond $R_h$ we see (Figure \ref{bb_2.25_1_3fig})
that the domain $D$ is now bounded by a periodic orbit, indicated by
$P$ in Figure \ref{bb_2.25_1_3fig}. Both arcs of $SM(X_{lb})$ now have the
edge character. The boundary of $B$, the basin of attraction of $O$,
is the union of $P$ with $SM(X_{lb})$. The left-hand arc of the
latter now coincides with the unstable manifold of the unstable
periodic orbit $P$.

\subsection{Hopf bifurcation}\label{hopf}
\begin{figure}
\centering \includegraphics[width=3.5in, height=2.5in]{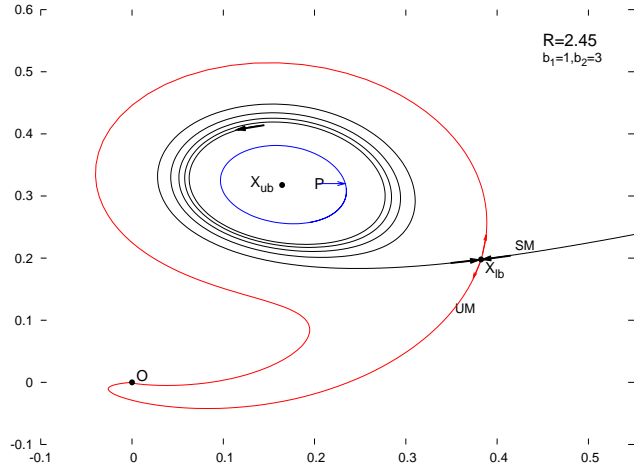}
\caption{\small The geometry of the phase portraits is qualitatively
the same as in Figure \ref{bb_2.25_1_3fig}. The value of $R$ is just less than $R_{bh}=2.5$, at which the
stability of $X_{ub}$ changes. The basin of attraction of $X_{ub}$ has
shrunk. The complementary region to the basin of attraction of $O$ now
consists only of this region (bounded by $P$) and the pair of curves
comprising $SM$, the stable manifold of $X_{lb}$.}\label{bb_2.45_1_3fig}
\end{figure}

\begin{figure}
\centering \includegraphics[width=3.5in, height=2.5in]{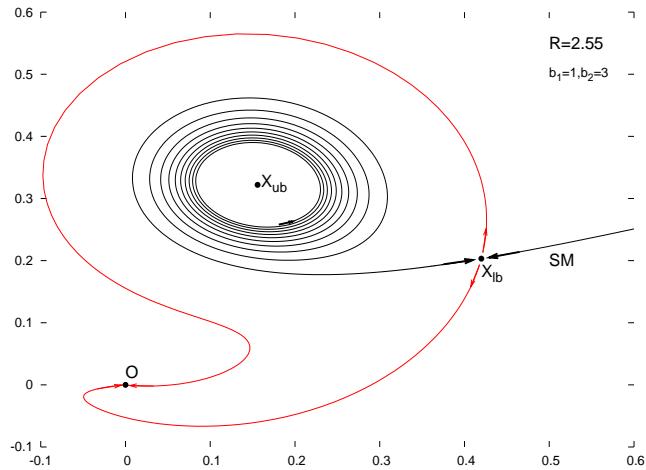}
\caption{\small In this figure, $R$ is just greater than
$R_{bh}=2.5$. The edge, consisting of the stable manifold of $X_{lb}$,
is now ``pure'' in the sense that all orbits except those lying on
this set relaminarize.}\label{bb_2.55_1_3fig}
\end{figure}

As $R$ is increased further, the basin of attraction of $X_{ub}$
begins to shrink (see Figure \ref{bb_2.45_1_3fig}) and the stability of that point weakens, i.e., the
(negative) real parts of the eigenvalues get smaller in absolute
value. This culminates in
a further bifurcation point $R=R_{bh}$\footnote{This is a 'backward
Hopf' bifurcation (thus $R_{bh}$), i.e., it would be a standard Hopf bifurcation if
we changed $t$ to $-t$ and ran $R$ backwards.} at
which the real parts of the eigenvalues vanish and the domain $D$
evanesces. For $R_{bh} < R < R_\infty$, essentially all of the ambient space lies in
$B$, i.e., essentially all orbits relaminarize. The only exceptions
are those lying on the stable manifold of $X_{lb}$, but these would be
sampled with probability zero (see Figure \ref{bb_2.55_1_3fig}). The
basin boundary is now a pure edge state.

\begin{figure}
\centering \includegraphics[width=3.5in, height=2.5in]{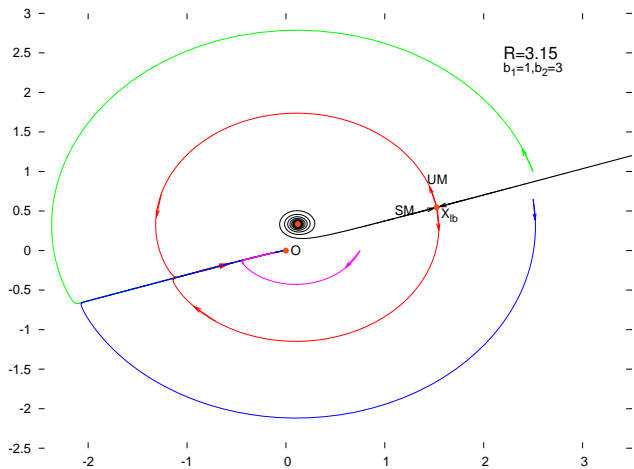}
\caption{\small Qualitatively like Figure \ref{bb_2.55_1_3fig} but for
value of $R$ approaching $R_\infty$, where $X_{lb} \to \infty$.
Three orbits are shown (green,blue,magenta) to indicate that most
orbits are similar to the arcs of $UM$.}\label{bb_3.15_1_3fig}
\end{figure}

\subsection{A final bifurcation}\label{infinity}

\begin{figure}
\centering \includegraphics[width=3.5in, height=2.5in]{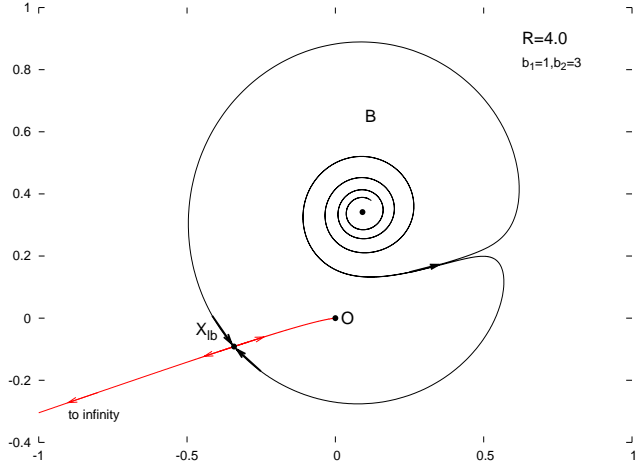}
\caption{\small The value of $R$ now exceeds that of a fourth critical
point $R_\infty=10/3$. The basin of attraction of the origin is now
finite. What appears to be a spiral surrounding the equilibrium point
$X_{ub}$ (unmarked in this figure) is in fact a pair of spirals very
close together. They are better resolved in Figure \ref{bb_5.0_1_3fig}
below.}\label{bb_4.0_1_3fig}
\end{figure}

\begin{figure}
\centering \includegraphics[width=3.5in, height=2.5in]{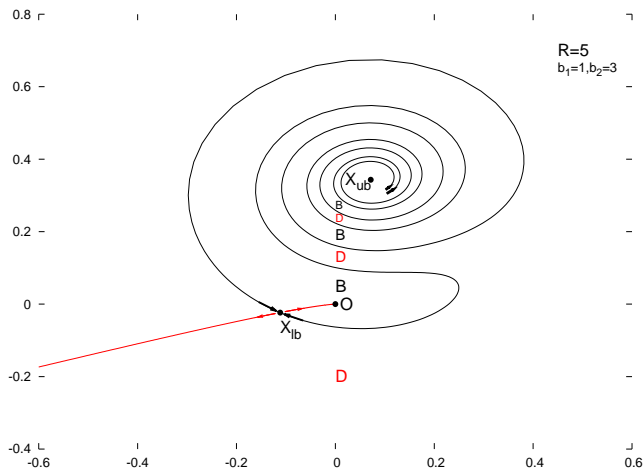}
\caption{\small Qualitatively similar to Figure
\ref{bb_4.0_1_3fig}. The regions marked $B$ and $D$ are respectively
the basin of attraction of $O$ and the region where all orbits
tend to $\infty$. The origin of coordinates now lies closer to the
basin boundary.}\label{bb_5.0_1_3fig}
\end{figure}

Finally, for $R>R_\infty$, the basin of attraction of the origin is
restored to finite size, as in Figure \ref{bb_4.0_1_3fig}, and the
simple picture of a basin is restored: the boundary separates orbits
that relaminarize from those that do not and there is no edge state. Moreover, as $R$ increases,
$X_{lb} \to O$ (like $1/R^2$ for large $R$) and therefore the distance from $O$ to $\partial B$
likewise tends to zero (like $1/R^3$; cf \cite{bt97} for this estimate). This is the basic picture of subcritical
transition, since perturbations of $O$ may find themselves outside $B$
despite being very small. Figure \ref{bb_5.0_1_3fig} is qualitatively
the same but for a larger value of $R$.

\section{Conclusions}\label{conclusions}
That certain features of transition in shear flows can be captured by
low-dimensional -- even two-dimensional -- models is well-known
(\cite{dama},\cite{ttrd},\cite{bdt},\cite{bt97}). The present study
adds to these features a picture of an edge state. It confirms the
view (\cite{skufca}) that the edge is a codimension-one invariant set
embedded in the basin of attraction of the laminar state. In the
present paper the edge state is in fact the stable manifold of the
unstable equilibrium point $X_{lb}$. It emerges as an edge state
simultaneously with the emergence of a periodic orbit via a homoclinic
bifurcation at $R=R_h$. This periodic orbit forms the boundary of the
basin of attraction of a further stable equilibrium point $X_{ub}$,
and the boundary $\partial B$ of the basin of attraction of the origin
consists of the union of this periodic orbit with the edge in the
interval $R_h < R < R_{bh}$.

A similar result appears in a four-dimensional model (\cite{nl}),
wherein an edge state likewise makes an appearance via a homoclinic
bifurcation. In that study, as in this one, simultaneously with the
appearance of the edge state there appears a periodic orbit $P$. In
that case, whereas the equilibrium point $X_{ub}$ is unstable, there
is a stable periodic orbit $Q$ with basin of attraction $D$. The
boundary $\partial D$ of $D$ contains the periodic orbit $P$,
which is a relative attractor: it is unstable but is an attractor for
orbits lying in $\partial D$. In \cite{nl} an interpretation of the
edge was offered as a pair of surfaces bounding an exquisitely narrow
gap containing points of $D$. However, in view of the results of the
present paper, it seems more likely that the edge, there also, is a
single surface: the stable manifold of the equilibrium point $X_{lb}$,
separating $B$ into two regions. In the nine-dimensional model of
\cite{skufca} the basin boundary similarly contains a chaotic,
invariant subset. In all these cases orbits starting near the basin
boundary may well participate in the dynamics of such invariant
subsets -- and therefore be quite complex -- for a while before
relaminarizing. In the higher-dimensional cases the complexity can be
heightened if the invariant sets in question are highly convoluted.

In the present model there is a final critical value $R_\infty$ of the
parameter beyond which the edge state has disappeared and the
geometrical structure of phase space is consistent with the
interpretation of a permanent, subcritical transition away from the
laminar state.  It is of course not clear that this result extends
beyond this simple model.

\bibliography{basinrev}
\end{document}